\documentclass{sig-alternate-05-2015}

\usepackage{verbatim}
\usepackage{hyperref}
\usepackage{graphicx}
\usepackage{enumitem}
\usepackage{color}
\usepackage{algorithm}
\usepackage{algorithmicx}
\usepackage{float}
\usepackage{algpseudocode}
\usepackage{multirow}
\usepackage{epstopdf}
\usepackage{tikz}
\usepackage{pgfplots}
\usetikzlibrary{arrows}
\usetikzlibrary{automata,positioning}
\usetikzlibrary{trees}
\usetikzlibrary{shapes}
\usepackage{xspace}
\usepackage{mathtools}
\hypersetup{
    colorlinks,
    citecolor=blue,
    filecolor=black,
    linkcolor=blue,
    urlcolor=blue,
    pdfauthor={Liangjie Hong and Adnan Boz},
    pdfsubject={Unbiased Data Collection and Exploration Strategy for PINT},
    pdftitle={Unbiased Data Collection and Exploration Strategy for PINT},
    pdfkeywords={unbiased data collection, exploration and exploitation, contextual multi-armed bandit}
}

\newdef{definition}{Definition}
\newfloat{algorithm}{t}{lop}

\makeatletter
\let\@copyrightspace\relax
\makeatother
\begin{document}


\title{An Unbiased Data Collection and Content Exploitation/Exploration Strategy for Personalization}

\numberofauthors{1}
\author{Liangjie Hong$^{\dag}$ and Adnan Boz$^{\S}$\\
\affaddr{$\dag$ Yahoo Research, Sunnyvale, CA, USA}\\
\affaddr{$\S$ Yahoo Personalization Team, Sunnyvale CA, USA}\\
\email{liangjie@yahoo-inc.com, adnanb@yahoo-inc.com}
}

\maketitle
\begin{abstract}
One of missions for personalization systems and recommender systems is to show content items according to users' personal interests. In order to achieve such goal, these systems are learning user interests over time and trying to present content items tailoring to user profiles. Recommending items according to users' preferences has been investigated extensively in the past few years, mainly thanks for the popularity of Netflix competition. In a real setting, users may be attracted by a subset of those items and interact with them, only leaving partial feedbacks to the system to learn in the next cycle, which leads to significant biases into systems and hence results in a situation where user engagement metrics cannot be improved over time. The problem is not just for one component of the system. The data collected from users is usually used in many different tasks, including learning ranking functions, building user profiles and constructing content classifiers. Once the data is biased, all these downstream use cases would be impacted as well. Therefore, it would be beneficial to gather unbiased data through user interactions. Traditionally, unbiased data collection is done through showing items uniformly sampling from the content pool. However, this simple scheme is not feasible as it risks user engagement metrics and it takes long time to gather user feedbacks. In this paper, we introduce a user-friendly unbiased data collection framework, by utilizing methods developed in the exploitation and exploration literature. We discuss how the framework is different from normal multi-armed bandit problems and why such method is needed. We layout a novel Thompson sampling for Bernoulli ranked-list to effectively balance user experiences and data collection. The proposed method is validated from a real bucket test and we show strong results comparing to old algorithms.
\end{abstract}

\section{Introduction}\label{sec:introduction}
It is critical for online service providers to show content items according to users' personal interests. The process is sometimes denoted as ``Personalization'' where user interests are learned over time and presentations of content items are tailored to user profiles. Recommending items according to users' preferences has been investigated extensively in the past few years, mainly thanks for the success of Netflix competition where the notion of ``Recommender Systems'' has been brought under the attention. 

A normal setting of a recommender system is that, when a user comes into the service, a few, ranging from a handful to hundreds, of items are shown to the user according to his/her profile. The user may be attracted by a subset of those items and interacts with them by clicking, commenting, re-posting, thumb-up or thumb-down those items. These user feedbacks are later incorporated into the next round of model training/testing/evaluation, in order to improve the effectiveness of personalization over time. In an ideal scenario, a user would go over all items prepared by the system and express his/her preferences. However, due to limited resource constraints like time budges and spaces, a user may only interact with very few, sometimes, even one or two items from the prepared list, leaving most items untouched. Therefore, systems are only learning from partial feedback data from users \cite{Rendle2009}. This cycle leads to significant biases into systems and hence results in a situation where user engagement metrics cannot be improved over time. The problem is not just for one component of the system. The data collected from users is usually used in many different tasks, including learning ranking functions, building user profiles and constructing content classifiers. Once the data is biased, all these downstream use cases would be impacted as well. Therefore, it would be beneficial to gather unbiased data through user interactions.

A generic approach to reduce biases is to show those under-represented items, balancing the set of items that are already interacted with users and the remaining set of items that are not interacted by or not even shown to users, with the risk of jeopardizing user experiences. The problem of exploiting and exploring items from the content pool with the aim to optimize a particular user engagement metric like Click-Through-Rate ({\tt CTR}) has been intensively studied in a wide range of scenarios (e.g., \cite{Agarwal2009a,Li2010,Tang2013}). In general, algorithms are proposed with a certain metric to optimize in mind. Each item contributes to the overall metric with uncertainty. As the exact amount of contribution is not known a prior, algorithms need to balance between choosing items with existing good performances and items with potential contributions. Different algorithms would utilize different strategies to exploit and explore items, either in deterministic or randomized ways.

Although existing exploit/explore frameworks are effectives tools to optimize a metric with uncertainty, they cannot be used directly to address the issue of biases and partial feedbacks from users. In particular, these methods are proposed to \textbf{optimize} a metric not to remove biases. Even if they are exploring items that are never shown before, once the uncertainty of these items with respect to contributing to the metric is exploited enough, algorithms would concentrate on the ones yielding largest rewards, achieving the optimal performance. On the contrary, any solution to reduce or remove biases in online services does not necessarily need to optimize any metrics. Indeed, for a mature online system, multiple, sometimes even competing, metrics might exist and thus, it is not wise for the solution to bias towards any one of them. Another extreme, which might mitigate the issue of biases, is to show items in uniformly random. Although this strategy will remove biases completely, it is not an efficient way to gather data from users. For instance, if a user faces a completely random items from a large content pool, it is very unlikely that these items are relevant to the user.

In this paper, we propose a novel unbiased data collection strategy and utilize Bayesian posterior sampling to balance exploitation and exploration to improve user experiences when gathering data. We demonstrate the effectiveness of the algorithm through a bucket test in France.

\section{Exploitation and Exploration}\label{sec:random}
In this section, we review a particular type of exploitation and exploration algorithms: contextual multi-armed bandit methods and we will develop them to suit unbiased data collection. In order to simplify the discussion, we use {\tt CTR} as the metric to optimize.
\subsection{Contextual Multi-Armed Bandit Problems}\label{sec:cmab}
A multi-armed bandit ({\tt MAB}) problem is a sequential decision making process. Bandit problems involve making a decision in each round. Once a decision is made an observation is collected and the corresponding reward computed. The contextual-bandit formalism generalizes this classic setting by introducing contextual information in the interaction loop.

Formally, we define by $\mathcal{A} = \{1,2,\cdots, K\}$, a set of \textit{actions}, a contextual vector $\mathbf{x}_{t} \in \mathcal{X}$, a \textit{reward} vector $\mathbf{r}_{t} = \{\mathbf{r}_{t,1}, \cdots, \mathbf{r}_{t,K}\}$, where each $\mathbf{r}_{t,a} \in [0,1]$ and a policy $\pi : \mathcal{X} \rightarrow \mathcal{A}$. A contextual bandit ({\tt cMAB}) describes a round-by-round interaction between a learner and the environment. At each round $t$, the problem can be decomposed into following steps:
\begin{itemize}
\item The environment chooses  $(\mathbf{x}_{t}, \mathbf{r}_{t})$ from some unknown distribution $\mathcal{D}$. Only $\mathbf{x}_{t}$ is revealed to the learner while the reward is not.
\item Upon observing $\mathbf{x}_{t}$, the learner uses some policy $\pi$ to choose an action $a \in \mathcal{A}$, and in return observes the corresponding reward $\mathbf{r}_{t,a}$.
\item (Optional) The policy $\pi$ is revised with the data collected for this round.
\end{itemize}Here, we define $\pi$ is parameterized by an unknown parameter $\theta$. Ideally, we would like to choose the policy maximizing the expected reward:
\begin{align*}
\arg\max_{\pi} \mathbb{E}_{\mathbf{x}, \mathbf{r} \sim \mathcal{D}} \Bigr[ \mathbf{r}_{\pi(\mathbf{x};\theta)}\Bigl]
\end{align*}If we are just interested in maximizing the immediate reward, then one should choose the action that maximize:
\begin{align*}
\int \mathbb{E}_{\mathbf{x}, \mathbf{r} \sim \mathcal{D}} \Bigr[ \mathbf{r}_{\pi(\mathbf{x};\theta)}\Bigl] P(\theta \, | \, \mathcal{Q}) \, d\theta
\end{align*}where $Q$ is the posterior distribution of $\theta$, given contextual information and reward information. As we discussed before, {\tt cMAB} would maximize expected reward, which may not be a good thing for reducing biases. However, {\tt cMAB} provides a nice framework to our later proposed method.

\subsection{Thompson Sampling for $K$-armed Bernoulli Bandit}
Thompson sampling \cite{Chapelle2011} is an effective way to conduct exploitation and exploration through Bayesian posterior sampling, optimizing {\tt CTR} in long-run. In an exploration/exploitation setting, we randomly selection an action according to its probability of being optimal. Therefore, we draw a set of random parameters $\theta$, which characterizes each arm in the current round, and pick the one that yields the maximum reward.

In the standard $K$-armed Bernoulli bandit, each action corresponds to the choice of an arm. The reward of the $i$-th arm follows a Bernoulli distribution with mean $\theta_{i}^{*}$. It is standard to model the mean reward of each arm using a Beta distribution since it is the conjugate distribution of the binomial distribution. The instantiation of Thompson sampling for the Bernoulli bandit is given in Algorithm \ref{algo:thompson_sampling}.
\begin{algorithm}
\caption{Thompson Sampling for the Bernoulli Bandit}
\label{algo:thompson_sampling}
\begin{algorithmic}
\State \textbf{Require:} $\alpha$, $\beta$ prior parameters of a Beta distribution
\State $S_{i} =0$ and $F_{i} =0$, $\forall i$ \{Success and failure counters\}
\For{$t =1, \cdots, T$}
	\For{$i=1, \cdots, K$}
  		\State Draw $\theta_{i}$ according to $\mbox{Beta}(S_{i}+ \alpha, F_{i}+\beta)$.
	\EndFor
	\State Draw arm $\hat{i} = \arg\max_{i} \theta_{i}$ and observe reward $r$
	\If{$r = 1$}
	  	  \State $S_{\hat{i}} = S_{\hat{i}} + 1$
	\Else
	  	  \State $F_{\hat{i}} = F_{\hat{i}} + 1$
	\EndIf
\EndFor
\end{algorithmic}
\end{algorithm}Although it seems promising in the first place, $K$-armed Bernoulli bandit can be hardly applied to real-world personalization ranking problems. One major issue of the algorithm is that it is an item-based strategy. Namely, in each round, only a single item is selected.

One way to extend an item-wise Bernoulli bandit to a list-wise Bernoulli bandit is to use a permutation of all items as an arm. Each permutation will be associated with a parameter and we could draw a sample from its posterior distribution to determine whether to choose this arm or not. However, it is in general difficult to estimate such parameters without any assumptions. With a strong independent assumption, one could sample the posterior distribution of a list-wise arm by jointly sampling posterior distributions from each individual item's posterior. With further assumptions, in a recently proposed work \cite{Gopalan2014}, the authors proposed such method to tackle the problem of playing subsets of bandit arms. The proposed method is to sample individual arms' parameters from their posterior distribution and then select the top $N$ arms deterministically. However, the optimality of the algorithm is not proven for the Beta/Bernoulli case. We treat it as a heuristic for Bernoulli ranked-list case, shown in Algorithm \ref{algo:bernoulli_rank_bandit}.
\begin{algorithm}
\caption{Thompson Sampling for Bernoulli Ranked-list Bandit}
\label{algo:bernoulli_rank_bandit}
\begin{algorithmic}
\State \textbf{Require:} $\alpha$, $\beta$ prior parameters of a Beta distribution
\State $S_{i} =0$ and $F_{i} =0$, $\forall i$ \{Success and failure counters\}
 \For{$t =1, \cdots, T$}
  	\For{$i=1, \cdots, K$}
  		\State Draw $\theta_{i}$ according to $\mbox{Beta}(S_{i}+ \alpha, F_{i}+\beta)$.
  	\EndFor
  	\State Sort $\theta_{i}$ in the descending order
  	\State Select the top $N$ items and observe $N$ rewards.
  	\State Update $S$ and $F$ for those $N$ items.
 \EndFor
\end{algorithmic}
\end{algorithm}

\section{Unbiased Offline Evaluation}\label{sec:causal inference}
In this section, we introduce the basic framework to conduct unbiased data collection. Before we dive into details, there are two basic design criterions any approach to such issue needs to meet:
\begin{enumerate}
\item The dataset is collected in an unbiased fashion or with bias but the bias could be countered in later analysis.
\item The proposed method needs to have a reasonable user engagement performance such that users do not need to suffer from any data collection strategies.
\end{enumerate}These two design criterions are usually not the focus in classic exploitation/exploration literature.

\subsection{Unbiased Data Collection Through {\tt cMAB}}
{\tt cMAB} is not only a powerful way to balance exploration and exploitation but also a method to construct unbiased offline evaluation process, suggested by \cite{Li2015}. The basic idea is that, we use a known policy to operate a {\tt cMAB} problem for collecting data and record the action to be performed, the reward, as well as the probability to select the winning arm at each round.

Similar to {\tt cMAB}, we have the following procedure:
\begin{enumerate}
\item The environment chooses  $(\mathbf{x}_{t}, \mathbf{r}_{t})$ from some unknown distribution $\mathcal{D}$. Only $\mathbf{x}_{t}$ is revealed to the learner while the reward is not.
\item Upon observing $\mathbf{x}_{t}$, the learner computes a multinomial distribution $\mathbf{p}$ over the actions $\mathcal{A}$. A \textbf{random} action $a$ is drawn according to the distribution and in return observes the corresponding reward $\mathbf{r}_{t,a}$.
\item The contexutal vector $\mathbf{x}_{t}$, the selected action $a$, reward $\mathbf{r}_{t,a}$ and the probability mass $\mathbf{p}_{a}$ are logged.
\end{enumerate}Comparing this to the standard {\tt cMAB} procedure, one could observe that the only difference is that, the optimal action is not pursued every single round while a random action is selected. As \cite{Li2014}, this is critical to perform causal inference in an offline environment and hence is important for unbiased data collection as well. Additionally, the probability to the winning action is logged where it can be used as propensity score for further analysis. In order evaluate the value of a policy $\pi$ \textit{offline}, the following estimator is used:
\begin{align}\label{eq:unbiased_offline}
\hat{V}(\pi) = \sum_{(x,a,\mathbf{r}_{a},\mathbf{p}_{a}) \in \mathcal{D}} \frac{\mathbf{r}_{a} \mathbb{I}(\pi(x) == a)}{\mathbf{p}_{a}}
\end{align}

This framework (but not exactly same) stemmed from \cite{Strehl2010}, also discussed in \cite{Li2011,Tang2013}. The main difference is that, these related work still uses the framework to evaluate {\tt cMAB} problems and therefore, requiring choosing the best possible arm in every round whereas in this framework, each arm is performed stochastically. Note that, if $\mathbf{p}$ is uniform over all arms, it is essentially uniformly random strategy that has been used frequently in the past, like \cite{Li2010,Chapelle2011}.

\subsection{Unbiased Policy Evaluation}\label{sec:unbiased_evaluation}
In this sub-section, we show that Equation \ref{eq:unbiased_offline} is an unbiased offline policy evaluation procedure. We define the value of a policy $\pi$ as:
\begin{align*}
V(\pi) &= \mathbb{E}_{(x,r) \in \mathbb{D}}\Bigr[ \mathbf{r}_{a} \Bigl] \\
&= \int_{(x,r) \in \mathbb{D}} \mathbf{r}_{a}  P(x,r) \, dx dr
\end{align*}In such case, we want to prove:
\begin{align*}
\mathbb{E}_{\mathcal{D} \in \mathbb{D}} \Bigr[ \hat{V}_{\mbox{offline}}(\pi)\Big] = V(\pi)
\end{align*}Let us expand the left hand side as:
\begin{align*}
\int_{\mathcal{D} \in \mathbb{D}} \Bigr[ \sum_{(x,a,\mathbf{r}_{a},\mathbf{p}_{a}) \in \mathcal{D}} \frac{\mathbf{r}_{a} \mathbb{I}(\pi(x) == a)}{\mathbf{p}_{a}} \Bigl] P(\mathcal{D}) \, d\mathcal{D}
\end{align*}The important step in the proof is that we need to make use of the following quantity:
\begin{align}\label{eq:random_arm}
\mathbb{E}_{a \in \mathbf{p}} \Bigr[ \frac{\mathbf{r}_{a} \mathbb{I}(\pi(x) == a)}{\mathbf{p}_{a}} \Bigl] = \mathbf{r}_{a} \mathbb{I}(\pi(x) == a) = \mathbf{r}_{\pi(x)}
\end{align}Thus, on expectation, we have:
\begin{align*}
& \int_{\mathcal{D} \in \mathbb{D}} \Bigr[ \sum_{(x,a,\mathbf{r}_{a},\mathbf{p}_{a}) \in \mathcal{D}} \frac{\mathbf{r}_{a} \mathbb{I}(\pi(x) == a)}{\mathbf{p}_{a}} \Bigl] P(\mathcal{D}) \, d\mathcal{D} = \\
& \int_{\mathcal{D} \in \mathbb{D}} \Bigr[ \sum_{(x, \mathbf{r}) \in \mathcal{D}} \mathbf{r}_{\pi(x)} \Bigl] P(\mathcal{D}) \, d\mathcal{D} = \int_{(x,r) \in \mathbb{D}} \mathbf{r}_{a}  P(x,r) \, dx dr
\end{align*}The last line comes from the fact that $(x,r)$ from $\mathcal{D}$ are sampled from $\mathbb{D}$ i.i.d. and $\mathcal{D}$ is a random sample from $\mathbb{D}$. The integral also mounts to $\mathbb{D}$. Note that the key part in Equation \ref{eq:random_arm} is that we need to select arms stochastically. 

\subsection{Thompson Sampling for $K$-armed Unbiased Data Collection}\label{sec:thompson_data_collection}
\begin{algorithm}
\caption{Thompson Sampling for Bernoulli Ranked-list Bandit}
\label{algo:bernoulli_rank_bandit_datacollection}
\begin{algorithmic}
\State \textbf{Require:} $\alpha$, $\beta$ prior parameters of a Beta distribution
\State $S_{i} =0$ and $F_{i} =0$, $\forall i$ \{Success and failure counters\}
\For{$t =1, \cdots, T$}
	\For{$i=1, \cdots, K$}
  		\State Draw $\theta_{i}$ according to $\mbox{Beta}(S_{i}+ \alpha, F_{i}+\beta)$.
  	\EndFor
  	\State \textbf{Compute $\mathbf{p}$ such that $p_{k} = \frac{\theta_{k}}{\sum \theta_{k}}$.}
  	\State \textbf{Sample $N$ items from $\mbox{Mult.}(\mathbf{p})$.}
  	\State Observe $N$ rewards $\mathbf{r}_{t}$.
  	\State Update $S$ and $F$ for those $N$ items according to $\mathbf{r}_{t}$.
  	\State Logging $N$ items, $\mathbf{p}$ and $\mathbf{r}_{t}$.
\EndFor
\end{algorithmic}
\end{algorithm}
We adapt Algorithm \ref{algo:bernoulli_rank_bandit} to the unbiased data collection case, shown in Algorithm \ref{algo:bernoulli_rank_bandit_datacollection}. The main difference between these two algorithms is that, in stead of deterministically choosing the best arm (ranked-list) in every round, Algorithm \ref{algo:bernoulli_rank_bandit_datacollection} chooses each arm probabilistically.

Algorithm \ref{algo:bernoulli_rank_bandit_datacollection} has several nice properties:
\begin{itemize}
\item It uses {\tt CTR} as an underlying metric to drive the data collection process. Therefore, the user engagement would be reasonable as high {\tt CTR} items will have higher chances to be selected into the ranked-list.
\item Both $\alpha$ and $\beta$ can be tweaked such that we control the variance of items to be sampled. Also, prior knowledge about items can be easily embedded into these parameters.
\end{itemize}Indeed, Beta distribution and Bernoulli bandits are not necessary for the algorithm. One can easily replace these settings with other underlying metrics and distributions. 

For implementation details, we notice that two steps, shown in bold in Algorithm \ref{algo:bernoulli_rank_bandit_datacollection}, are expensive for every request, give we could have millions of items. In particular, it is slow to sample $N$ out of $M$ items without replacement for every round. Here, we use a further approximation. Instead of sampling $N$ from $M$ items, with proportional to their probability to be clicked, we compute the following quantity:
\begin{align*}
\theta_{i}^{\prime} = \lambda \theta_{i}
\end{align*}where $\lambda \sim \mbox{Unif.}(0,1)$. Then, we sort $\theta_{i}^{\prime}$ and select the top $N$ items. This procedure is much more efficient and also supported as:
\begin{align*}
\mathbb{E}[P(c \, | \, i)] = \mathbb{E}[\theta_{i}^{\prime}] = \frac{1}{2} \mathbb{E}[\theta_{i}]
\end{align*}Therefore, the expectation to be clicked is within a constant factor of the expectation of $\theta_{i}$. Note that, we do need to compute the normalization factor by using $\theta_{i}^{\prime}$ values.

\section{Online Experiments}\label{sec:online_buckets}
\begin{table*}
\centering
\caption{Comparison between the old exploration algorithm and the new exploration algorithm on a number of key distributions.}
\begin{tabular}{l||c|l|l|l}
\textbf{Algorithm} & \textbf{Metrics} & \textbf{Skewness} & \textbf{Mean} & \textbf{Median} \\ \hline \hline
New Algorithm & \multirow{2}{*}{View Distribution} & $6.76$ & $10,868.46$ & $2,500.00$ \\
Old Algorithm & & $9.65$ & $2,328.70$ & $441.50$ \\ \hline
New Algorithm & \multirow{2}{*}{Click Distribution} & $14.46$ & $1,059.25$ & $64.00$ \\
Old Algorithm & & $14.64$ & $241.17$ & $7.00$ \\ \hline
New Algorithm & \multirow{2}{*}{CTR Distribution} & $2.28$ & $0.04$ & $0.03$ \\
Old Algorithm & & $3.87$ & $0.03$ & $0.02$ \\ \hline
New Algorithm & \multirow{2}{*}{Item Cold-Start Distribution} & $1.15$ & $37.26$ & $13.86$ \\
Old Algorithm & & $3.47$ & $100.02$ & $13.05$ \\ \hline
\end{tabular}
\label{table:comparison}
\end{table*}
\begin{figure}
\centering
\includegraphics[scale=0.35]{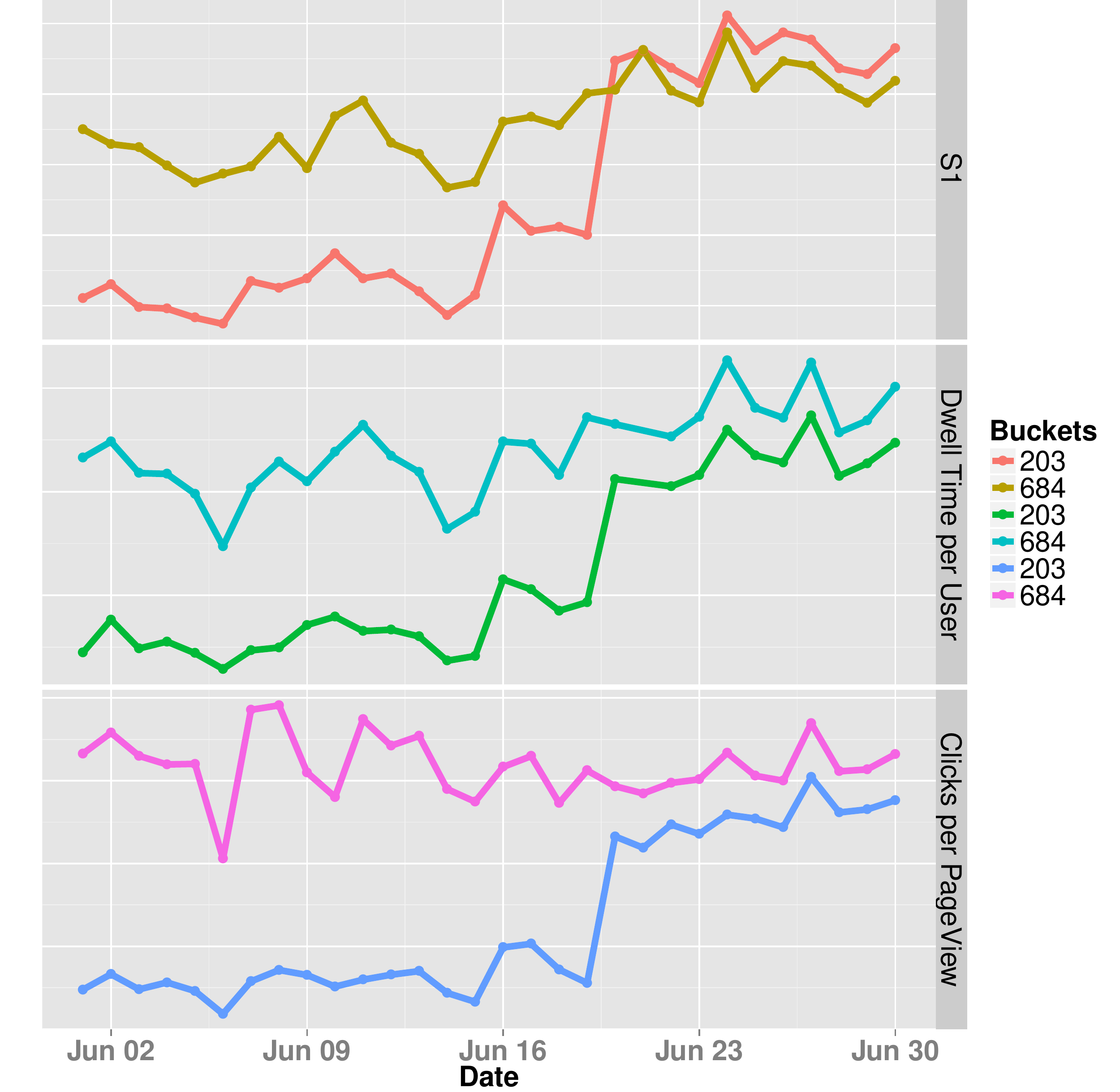}
\caption{User Engagement Metrics Before/After Launching New Algorithm.}
\label{fig:user_metrics}
\end{figure}
In this section, we demonstrate how one version of Algorithm \ref{algo:bernoulli_rank_bandit_datacollection} is deployed in production and its effectiveness from a bucket test. We use Yahoo's France Homepage (Slingstone France) as a testbed. Before we deployed this algorithm, Slingstone France was running a deterministic ranking which is essentially to rank items chronologically. As we will see, this algorithm has a poor user engagement metric. For using our proposed algorithm, we use $6$-hour average {\tt CTR} for a bucket (5\% of traffic) as the parameter for $\alpha$ and $\beta$. Therefore, a new item would have a reasonable starting point, instead of zero {\tt CTR}. We launched the algorithm in bucket $203$ in France in June 20, 2014. For comparison purpose, we compare the bucket to a General-Available (GA) bucket (bucket 684), which has the same size of traffic. Here, we are interested in three user engagement metrics:
\begin{itemize}
\item \textbf{Dwell Time Per Depth}: Post-click dwell time on article pages, divided by the total number of items viewed in the stream, denoted as S1.
\item \textbf{Dwell Time Per User}: Total amount of dwell time, divided by the total number of users.
\item \textbf{Clicks per PageView}: Total number of clicks per pageview.
\end{itemize}We show metric changes before and after the launch of the new algorithm in Figure \ref{fig:user_metrics}. We are interested in two things: 1) user engagement metrics should be improved by the new algorithm, 2) the new algorithm is not optimizing {\tt CTR} as well. From the figure, we can obviously see that all three user engagement metrics has been improved since June 20th. Before that, bucket 203, which was running old deterministic chronological stream, suffered from all metrics. Right after the launch, all metrics jumped closer to the GA bucket. However, we do notice that both {\tt Dwell Time Per User} and {\tt Clicks per PageView}, two main indicators for {\tt CTR}, do not perform as good as GA, meaning that the algorithm is \textbf{optimizing} {\tt CTR}, which is a good thing for the data collection bucket. At the same time, S1 is always comparable as the GA bucket, implying that users engage with the stream in the data collection bucket as well.

For the data collection bucket, we are also interested in how it is effectively exploring the whole content pool. In other words, the data collection bucket should pay more attention to all kinds of items and the skewness of its view distribution, click distribution should be less than the previous algorithm. We show the comparison between the old algorithm and the new algorithm in Table \ref{table:comparison} in a number of distributions. We can see that, from view distributions, click distributions and {\tt CTR} distributions, the newly proposed algorithm significantly improves the skewness and the distributions are more balanced. Indeed, the results show that the view skewness improved 30\%, click skewness improved 2\%, {\tt CTR} skewness improved 30\% and the article cold-start has improved dramatically; the articles seen in the first $30$ minutes went up from 66\% to 81\% and in the first 2 hours went up from 82\% to 92\%.

\section{Conclusion and Future Work}\label{sec:conclusion}
In this paper, we introduce a new algorithm to gather unbiased data with reasonable user engagement metrics. We discussed how it differs from traditional exploitation and exploration work and why the proposed framework would gather unbiased data. We demonstrated the effectiveness of the proposed approach through a live bucket test and showed that the method significantly improved the user engagement metrics and the skewness of a number of distributions of items.For future work, we would like to compare more variants of the framework and train machine learning models (e.g., ranking models, user profiling models) from the data we gathered to demonstrate that model training process can benefit from the data.

\bibliographystyle{abbrv}
\bibliography{source}
\end{document}